\documentclass[a4paper]{aa}

\usepackage{graphicx}
\usepackage{txfonts}
\usepackage{natbib}
\bibpunct{(}{)}{;}{a}{}{,}

\newcommand{\gr}{$\gamma$-ray}
\newcommand{\grs}{$\gamma$-rays}

\newcommand{\dg}{\ensuremath{^\circ}}
\newcommand{\hess}{H.E.S.S.}
\newcommand{\mrm}{\mathrm}

\begin{document}

  \title{Background Modelling in Very-High-Energy \gr\ Astronomy}

  \author{D.~Berge\inst{1,2}
    \and S.~Funk\inst{1,3}
    \and J.~Hinton\inst{1,4,5}}

  \offprints{D. Berge, \email{berge@cern.ch}}

  \institute{
    Max-Planck-Institut f\"ur Kernphysik, P.O. Box 103980, D-69029
    Heidelberg, Germany
    \and
    CERN, CH-1211 Geneva 23, Switzerland
    \and
    Kavli Institute for Particle Astrophysics and Cosmology, Stanford
    University, 2575 Sand Hill Road, Menlo Park, CA 94025, USA
    \and
    Landessternwarte, K\"onigstuhl, D 69117 Heidelberg, Germany
    \and
    School of Physics \& Astronomy, University of Leeds, Leeds LS2 9JT, UK
  }
  \date{Received 31 October 2006 / Accepted xx October 2006}

  \abstract
      {Ground based Cherenkov telescope systems measure astrophysical
	\gr\ emission against a background of cosmic-ray
	induced air showers.  The subtraction of this background is a
	major challenge for the extraction of spectra and
	morphology of \gr\ sources.
      }
      {The unprecedented sensitivity of the new generation of ground
	based very-high-energy \gr\ experiments such as \hess\ has lead
	to the discovery of many previously unknown extended
	sources. The analysis of such sources requires a range of
	different background modelling techniques. Here we describe
	some of the techniques that have been applied to data from the
	\hess\ instrument and compare their performance.}
      {Each background model is introduced and discussed in terms of
	suitability for image generation or spectral
	analysis and possible caveats are mentioned.}
      {We show that there is not a single multi-purpose model,
	different models are appropriate for different tasks. To keep
	systematic uncertainties under control it is important to
	apply several models to the same data set and compare the
	results.} 
      {}

  \authorrunning{Berge, Funk, Hinton}
  \titlerunning{Background modelling in \gr\ astronomy}
  
  
  \maketitle
  
  \section{Introduction}
  \label{sec:intro}
  
  Ground based very-high-energy \gr\ telescope systems such as
  \hess~\citep{HESS}, MAGIC~\citep{MAGIC}, VERITAS~\citep{VERITAS} and
  CANGAROO-3~\citep{CANGAROO3} have greatly increased the sensitivity
  of the Atmospheric Cherenkov technique. However, these instruments
  can only reach their full potential if systematic effects are
  brought fully under control.  A major challenge for
  experiments of this type is the subtraction of the background of
  non-\gr\ induced air showers. This background can be
  dramatically reduced using image-shape selection criteria, but
  cannot be removed completely. The background above a few hundred GeV
  is dominated by hadronic cosmic-ray showers, with
  cosmic-ray electrons increasingly important at low energies and
  after tight image selection cuts. The subtraction of this background
  is the main source of systematic errors and, if not done correctly, 
  can even produce an artificial source.

  For single telescope instruments (for example the pioneering Whipple
  telescope~\citep{WHIPPLE}) the classical approach to background
  subtraction was the \emph{ON/OFF} observing mode. In this mode
  observations (\emph{runs}) centred on the target source are
  interspersed with equal--length observations of an empty field at
  equal zenith angle (typically a region offset in Right Ascension by
  30~minutes). The background is assumed to be equal in the two runs,
  the difference between them provides a measurement of the
  \gr\ signal. A major drawback of this approach is that only
  half of the available dark time is spent \emph{ON-source}. The
  \emph{wobble-mode} pioneered by the HEGRA
  collaboration~\citep{HEGRA} avoids this problem by keeping the
  targeted source region in the field of view (\emph{FoV}) at all
  times, with an alternating offset relative to the system's
  pointing direction (for point sources typically $\pm 0.5\dg$ in
  Declination). A background estimate (\emph{OFF} data) for the source
  region (\emph{ON} data) can then be derived from a region on the
  opposite side of the FoV from the same run as the \emph{ON} data.

  \begin{figure*}
    \centering
    \includegraphics[width=16cm,draft=false]{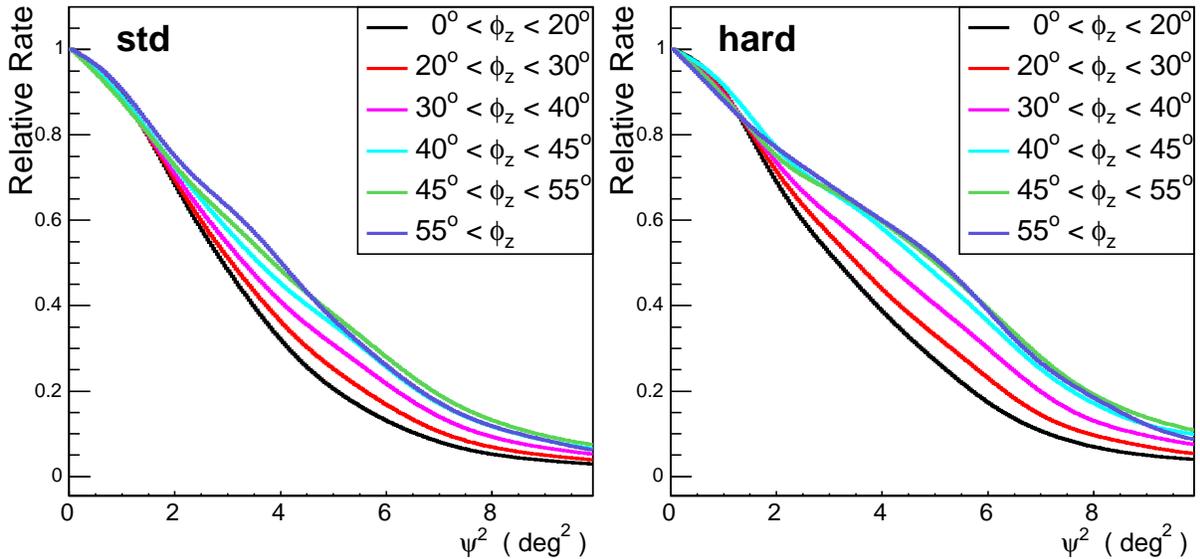}
    \caption{
      The variation of the radial system
      acceptance function with zenith angle for two different sets of
      cuts. The distributions show the squared angular distance
      $\psi^2$ between reconstructed event directions and pointing
      direction of the telescope system generated from
      \emph{OFF}-source data. The curves are generated by smoothing
      the one-dimensional acceptance histograms. They are arbitrarily
      normalised to 1 at the system pointing direction, corresponding
      to $\psi^2=0$. \textbf{Left:} The standard loose set of cuts
      (labelled \textit{std}) is shown. It is mainly used for the
      determination of \gr-ray spectra. \textbf{Right:} The
      \textit{hard} cuts employing a larger cut on the minimum image
      amplitude are shown. This configuration is typically used for
      morphology studies.}
    \label{acc_shape}
  \end{figure*}

  For wide FoV instruments the probability of serendipitous detection
  of non-targeted sources rises dramatically (particularly for
  observations close to the Galactic plane). The discovery of such
  sources (for example
  HESS\,J1303--631~\citep{HESSJ1303}) demonstrates the need for
  background models that provide background estimates for the whole
  FoV of the instrument. Any systematic survey of a whole sky region
  also requires such models.  The importance of surveys has been
  demonstrated by the recent discovery of more than ten new
  \gr\ sources in a survey of the inner Galaxy with
  \hess~\citep{HESSSCAN,HESSSCAN2}. Moreover, extended \gr\ sources
  such as RX~J1713.7--3946~\citep{HESSRXJ1713_I,HESSRXJ1713_II} and
  RX~J0852.0--4622 (Vela~Junior)~\citep{HESSVelaJnr} present
  additional difficulties for background subtraction. They require
  correct background modelling over a region of the sky that is
  substantially larger than the source itself.

  Many different approaches to background modelling are possible and
  have been applied in the analysis of data from the \hess\
  instrument. These models have different strengths and weaknesses and
  it is usually desirable to apply several of them in the analysis of
  any \gr\ source to get a handle on the systematic
  uncertainties connected to the background determination. In this
  paper we aim to describe some of these approaches.  While we will be
  using \hess\ data as the basis for our examples it should be noted
  that the techniques described here apply generally to VHE
  \gr\ instruments. Where necessary we will point out
  features that are specific to the analysis applied or to
  \hess\ and its large FoV.

  An outline of this paper is as follows: We begin in
  section~\ref{sec:models} with a general introduction of background
  modelling and investigate in detail the properties of the
  \emph{system acceptance}, as it is a key starting point for almost
  all background models. We then move on to a description of
  individual background models. In section~\ref{sec:comparison} we
  compare the results of different approaches applied to
  particular data examples. In section~\ref{sec:stars},
  the impact of bright stars on observations using atmospheric
  Cherenkov telescopes is demonstrated. Section~\ref{sec:choice}
  summarises the strengths and weaknesses of the different background
  models and describes to which purpose they are best suited.

  \section{Background Modelling}
  \label{sec:models}
  With cuts on image-shape parameters the cosmic-ray background can be
  reduced by a factor of $\sim$100, resulting in a
  signal-to-background ratio for a strong point source like the Crab
  nebula on the order of $1\mrm{:}1$~\citep{HESSCrab}. The remaining
  background of \gr-like events must be estimated to derive the
  significance of any possible \gr\ signal. Given a number of counts
  $N_{\mathrm{on}}$ in a test region, and $N_{\mathrm{off}}$ counts in
  a background control region, the \gr\ excess is defined as
  \begin{equation}
    \label{eq:excess}
    N_{\mathrm{excess}} =
    N_{\mathrm{on}} - \alpha\,N_{\mathrm{off}} \, . 
  \end{equation}
  The parameter $\alpha$ is a normalisation factor which accounts for
  solid-angle, exposure-time, zenith-angle, and acceptance differences
  between the test region and the background control region. In
  general, it is the ratio of the effective (acceptance-weighted)
  exposure integrated in time and angular space over the signal and
  background region, usually referred to as the \emph{ON} and
  \emph{OFF} regions. It can generally be defined as:
  
  \begin{equation}
    \label{eq:alpha}
    \alpha = \frac{\int_\mrm{on} A^\gamma_\mrm{on}
      (\theta_x,\theta_y,\phi_{\mrm{z}},t) \, \mrm{d} \theta_x \, \mrm{d}
      \theta_y \, \mrm{d} \phi_{\mrm{z}} \, \mrm{d} t}{\int_\mrm{off}
      A^\gamma_\mrm{off} (\theta_x,\theta_y,\phi_{\mrm{z}},t) \,
      \mrm{d} \theta_x \, \mrm{d} \theta_y \, \mrm{d} \phi_{\mrm{z}}
      \, \mrm{d} t}
    \, .
  \end{equation}
  $A^\gamma_\mrm{on,off}$ is the system acceptance of \gr\ like events
  and depends on the position $(\theta_x,\theta_y)$ in the FoV and the
  zenith angle $\phi_{\mrm{z}}$ of observations. Additionally,
  different exposure times $t$ for ON and OFF region have to be taken
  into account. Note that if ON and OFF region are different in size
  and shape, the integration of the system acceptance in
  Eq.~\ref{eq:alpha} must take this into account. Given a number of
  \emph{ON} and \emph{OFF} counts and $\alpha$, the statistical
  significance ($S$) of the excess is typically calculated following
  the prescription of \citet[][ equation~17]{LiMa}.

  The task of a background model is to provide the quantities
  $N_{\mathrm{off}}$ and $\alpha$. A choice of background regions such
  that $\alpha\ll 1$ (achievable, e.g., by choosing much larger OFF
  than ON regions) results generally in higher statistical
  significance, because background fluctuations are reduced, but may
  also result in increased systematic errors. The principle difficulty
  in deriving a background estimate is the determination of the
  correct value of $\alpha$.
  Since proper control over the system acceptance is crucial for this purpose,
  we investigate below the acceptance of \hess.


  \subsection{Cosmic-Ray System Acceptance after \gr\ Cuts}
  \label{subsec:acceptance}
  The system acceptance is defined as the probability of accepting,
  after triggering, analysis cuts and \gr\ selection,
  a background event reconstructed at a certain position in the system 
  FoV and with a certain energy.   
  For most background models some knowledge of the system acceptance
  is required to generate an image of \gr\ excess events or calculate
  significances of arbitrary positions in the FoV. In general, the
  acceptance depends on:
  \begin{itemize}
  \item[$\Box$] the position in the FoV, particularly the distance to 
   the optical axis
  \item[$\Box$] the zenith ($\phi_{\mrm{z}}$) and azimuth 
   $\vartheta_{\mrm{az}}$ angle of observations, due to the influence of the Earth's magnetic 
   field on the shower development in the atmosphere and the 
   rotational asymmetry of the telescope system.
  \item[$\Box$] the reconstructed primary energy
  \item[$\Box$] the time of observation, due to possible changes in the system
   configuration and aging of mirrors 
  \item[$\Box$] the sky coordinates viewed, due to the night-sky
   background light level.
  \end{itemize}
  In most cases it is a reasonable assumption that the acceptance is
  radially symmetric (the validity of this assumption is discussed
  later).  It is generated in a one-dimensional fashion as the number
  of background events as a function of the (squared) angular distance between
  reconstructed event direction and system pointing direction. It can
  either be determined on a run-by-run basis from the data set under
  analysis or be extracted from observations without significant \gr\
  emission in the FoV (\emph{OFF} runs). In the latter case it is
  assumed that the system acceptance is identical for the \emph{ON}
  and \emph{OFF} runs. In the former case one may face two problems,
  \gr\ contamination by a source, and lack of statistics. For a typical
  data run lasting 28~minutes, recorded at moderate zenith angles,
  the number of events after \gr\ selection cuts (available for the
  determination of the acceptance) is $\mathcal{O}(10^4)$ with cuts
  for spectral analysis, and as low as $\mathcal{O}(10^3)$ with cuts
  for morphology studies (see below for a description of the analysis
  cuts). 

  Here we use 220 hours of \hess\ observations without significant 
  \gr\ sources in
  the FoV to obtain a model of the radial system acceptance. These
  reference observations are sub-divided into zenith-angle
  bands. Events passing \gr\ cuts (i.e. \gr-like background events)
  are then binned according to the squared angular distance between
  the reconstructed event direction and the system's pointing
  direction. Figure~\ref{acc_shape} illustrates the dependence of the
  acceptance on the zenith angle of observations and analysis
  cuts. The
  shallow central peak and rapid decline towards larger distances,
  stems from the analysis based on image (\emph{Hillas}) parameters,
  where a cut on the \emph{distance} between image centre-of-gravity
  and the camera centre is applied to avoid truncation effects at the
  camera edge. Due to the finite camera size,
  edge effects are inherent and will always appear independent of the
  exact analysis applied.

  \begin{figure*}
    \centering
    \includegraphics[width=16cm,draft=false]{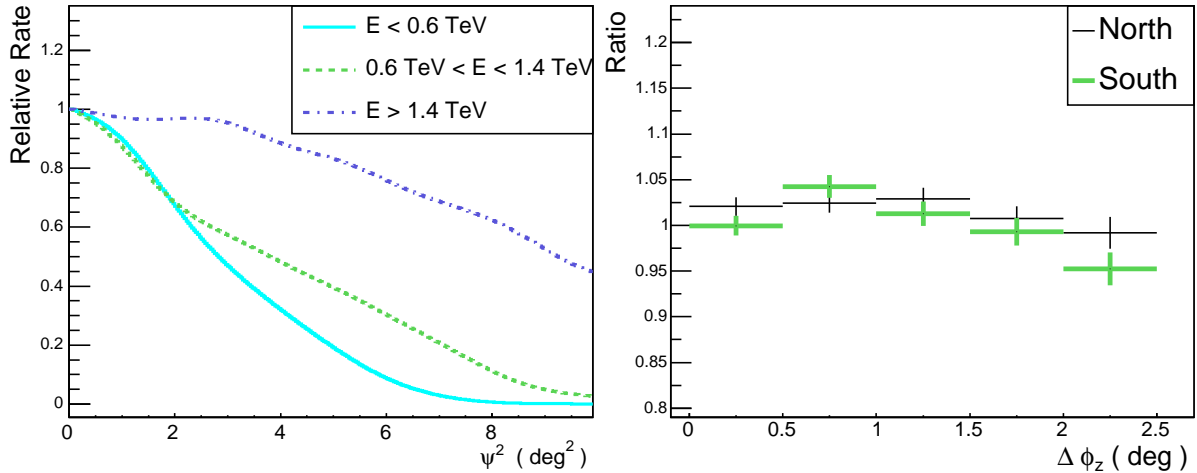}
    \caption{\textbf{Left:} The energy dependence of the system
      acceptance is demonstrated. For the three different energy bands
      shown, the shape of the acceptance broadens dramatically with
      increasing energy. \textbf{Right:} Plot to test for a
      zenith-angle dependent FoV gradient (note the change of scales
      compared to the left-hand side). For \emph{OFF} data taken at
      zenith angles of $40^{\circ}$ to $45^{\circ}$, the symmetry of
      the system acceptance along an axis running parallel to the
      zenith direction is investigated. For this purpose we fill all
      events into a histogram in bins of zenith-angle difference
      $\Delta \phi_\mrm{z}$ between the event direction and the system
      pointing direction. To test the symmetry with respect to the
      pointing direction, the ratio is calculated from the number of
      events in bins on either side of the central
      $\Delta\phi_\mrm{z}=0$ bin, equally far away from this bin. In
      the plot we show this ratio versus $|\Delta\phi_\mrm{z}|$. The
      data are split according to the azimuth of the observations: to
      the north (black crosses) and to the south (green crosses). Note
      that all corrections accounting for mis-pointing of the
      telescopes and for zenith-angle dependent trigger rates (see
      main text) are included here.}
    \label{acc_energy}
  \end{figure*}

  As can be seen from Fig.~\ref{acc_shape}, 2\dg\ away from the system centre, 
  the \gr\ acceptance at moderate
  zenith angles decreases to 20\% - 50\% of the peak value, depending
  on analysis cuts. 
  In addition a smooth variation with zenith angle is
  apparent. With increasing zenith angle, the system acceptance
  broadens, an increasing fraction of events with directions further
  away from the system pointing direction is detected. This is a
  direct consequence of the fact that with increasing zenith angle the
  shower maximum is increasingly further away from the telescope
  system causing a broadening of the Cherenkov light-pool on ground
  and hence an enlarged phase space for events with large inclination
  angles. When comparing the average curve for any given zenith-angle
  band to the radial acceptance in different fields of view, observed
  at the same altitude, the scatter is relatively small, less than 3\%
  within 1\dg\ of the observation position and less than 10\% out to
  3\dg. It is therefore justifiable to use \emph{OFF} data taken in
  different fields of view to determine a model of the system
  acceptance.

  The influence of analysis cuts is also apparent in
  Fig.~\ref{acc_shape}. The two sets of cuts used throughout this
  paper are labelled \emph{std} and \emph{hard}.  The first set
  includes a cut on the minimum amplitude of each camera image at
  80~photo-electrons (p.e.) and is optimised for the determination of
  source spectra. The second set uses a cut on the image amplitude at
  200~p.e., and provides better background suppression and superior
  angular resolution. 
  It is therefore
  normally used for source searches and image generation (more
  detailed descriptions of the \hess\ analysis techniques may be found
  in \citet{HESSPKS2155} and \citet{HESSRXJ1713_II}). The larger cut
  on the minimum image size results in curves which exhibit a generally
  less pronounced peak and a less rapid decline towards large distances.
  There is an increased fraction of events with large
  inclination angles with respect to the system pointing direction.
  
  The azimuth dependence of the radial system acceptance is small and
  therefore neglected here: when sub-dividing data taken in a narrow
  zenith-angle band into azimuth bins (say North, East, South, and
  West), only marginal differences occur at the few-percent level. The
  energy dependence of the acceptance is much stronger, greatly
  complicating the use of background models that require an acceptance
  correction for spectral analysis. This is illustrated in
  Fig.~\ref{acc_energy}~(left) where the energy dependence for a
  zenith angle range from 0\dg\ to 20\dg\ is plotted. The curves shown
  correspond to three different energy bands, $E < 0.6~\mrm{TeV}$ ,
  $0.6~\mrm{TeV} < E < 1.4~\mrm{TeV}$, and $1.4~\mrm{TeV} < E$. For
  relatively small energies the acceptance declines rapidly with
  increasing offset. For large energies the shape is completely
  different. High-energy showers result in large Cherenkov light-pool
  radii on ground. Therefore, as already mentioned, more events with
  large angular offsets start to trigger the array. In effect, for
  energies beyond 1.4~TeV, the acceptance is almost flat out to a
  distance of 2\dg\ from the system pointing direction.
  
  As previously mentioned, in most cases the system acceptance is
  assumed to be radially symmetric. The most intuitive cause of
  deviations from radial symmetry is a zenith-angle dependent linear
  gradient across the FoV. The larger the zenith angle of observation,
  the larger the effective energy threshold of the system, due to the
  increasing absorption of showers. Since the energy spectrum of the
  cosmic-ray background is rather steep, the trigger rate, and thus
  the event rate, of the system decreases smoothly with increasing
  zenith angle~\citep{FunkTriggerPaper}. Hence, in the \hess\ FoV of
  $\approx 5\dg$ significant variations of the system acceptance along
  the zenith axis may occur, and indeed such variations are
  observed. Depending on zenith angle the peak system acceptance
  extracted from \textit{OFF} runs does not coincide with the nominal
  centre of the system pointing direction. It is found to be slightly
  shifted towards smaller zenith angles, away from the pointing
  direction. To account for this effect, the zenith-angle dependence
  of the shift is determined: for small zenith angles $(\sim 10\degr)$
  the shift is negligible $(< 0.01\degr)$, at moderate $30\degr$ it is
  on the order of $0.03\degr$ and exceeds $0.13\degr$ for very large
  zenith angles beyond $60\degr$. A parametrisation is used to correct
  the nominal centre of the system pointing direction in each
  \emph{OFF} run. Figure~\ref{acc_energy}~(right) explores remaining
  deviations from radial symmetry along the zenith axis. For this
  purpose, \emph{OFF} runs are processed in zenith-angle bands,
  storing event distributions as a function of the zenith-angle
  difference between the pointing direction of the system and the
  event direction $\Delta \phi_\mrm{z}$. From the resulting
  distributions (ranging from $\Delta \phi_\mrm{z} = \pm 3.5\dg$) a
  ratio is created dividing the number of events in a certain bin on
  the positive $\Delta \phi_\mrm{z}$-side by the number of events in
  the corresponding bin on the negative side to test for symmetry
  around the zenith angle of observation. The resulting distribution
  is shown in Fig.~\ref{acc_energy}~(right) for two azimuthal bands
  (north and south) to search for effects related to the Earth's
  magnetic field. If there was no zenith-angle dependence, the ratio
  would be 1 for the whole FoV. There seems to be a residual
  distortion of the system acceptance along the zenith axis, in the
  direction one would expect from the trigger-rate variation. Larger
  zenith angles have smaller event numbers, indicating a slight
  under-correction of this effect. However, remaining deviations are
  estimated to be less than 5\%. Within statistics, there is no
  North-South effect apparent, the event-ratio distributions are in
  reasonable agreement with each other.

  \begin{figure*}
    \centering
    \includegraphics[width=14cm,draft=false]{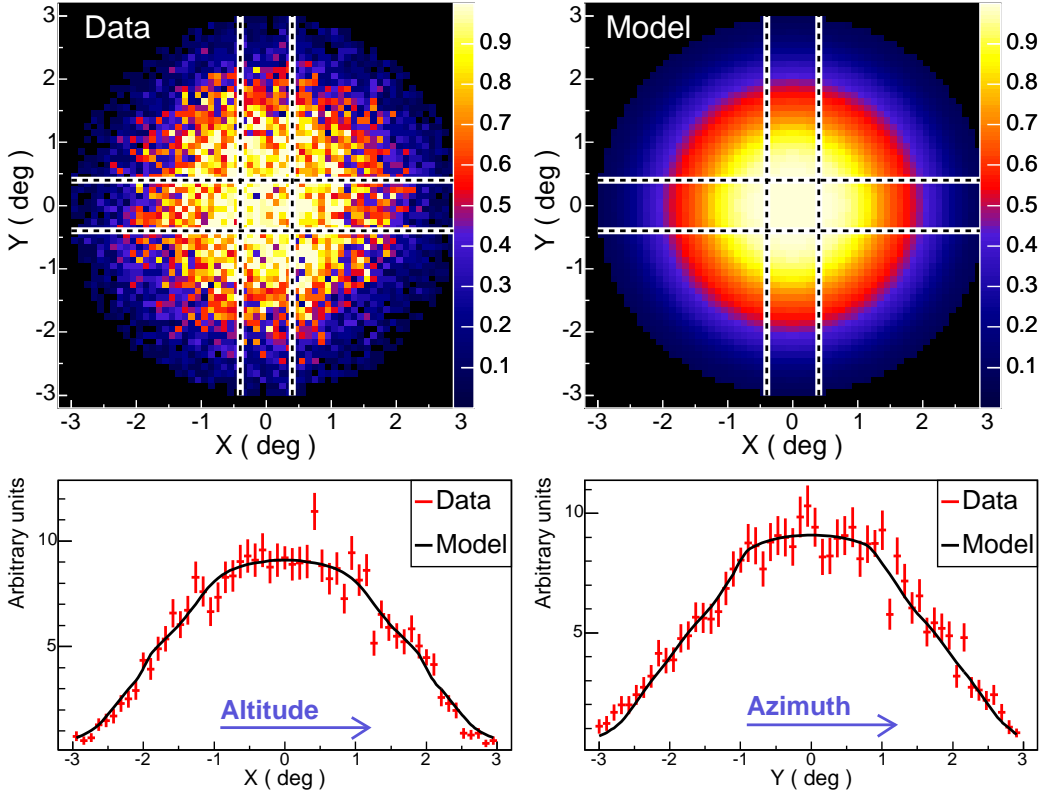}
    \caption{Plots of the system acceptance are shown for observations
      of the supernova remnant SN~1006, compared to the radially
      symmetric model acceptance determined from \emph{OFF}
      runs. \textbf{Upper panel:} Plots labelled ``Data'' and
      ``Model'' are the (arbitrarily normalised) acceptances
      determined in the \emph{nominal system} on a run-by-run
      basis. Positive $x$-direction corresponds to positive altitude,
      positive $y$-direction to positive azimuth. The parallel
      horizontal and vertical lines define bands ($0.8\dg$ wide) used
      to produce slices for a one-dimensional comparison, shown in the
      \textbf{lower panel:} The two plots show projections along $x$
      and $y$ through data (red crosses) and model map (black lines)
      within the thick bands indicated in the upper panel.}
      \label{acc_2d_symmetry}
  \end{figure*}

  The validity of the simplifying assumption that the system
  acceptance is radially symmetric can be verified with data. For that
  purpose such a symmetric model acceptance derived from \emph{OFF}
  data can be compared to the acceptance of a single data set without
  \gr\ source, for example the \hess\ data from the 2003-2004
  observation campaign of SN~1006~\citep{HESSSN1006} (which shows no
  evidence of \gr\ emission). Reconstructed directions of \gr-like
  events are plotted in a coordinate system centred on the system
  pointing direction in the ``altitude-azimuth'' system (the system is
  also referred to as \textit{nominal system}). Accumulating events
  from different data runs one obtains in this way a sample
  two-dimensional \gr\ acceptance map. This can be compared to a model
  acceptance map derived by choosing, for each run, the
  one-dimensional radial acceptance for the corresponding zenith angle
  (cf.\ Fig.~\ref{acc_shape}). The acceptance is then rotated in the
  nominal system and accumulated for all runs yielding an acceptance
  model which can be compared to the system acceptance deduced
  from the data set. The result, derived from 6.3 hours (after
  dead-time correction) of (4-telescope) \hess\ observations of
  SN~1006, is shown in Fig.~\ref{acc_2d_symmetry}. There is general
  agreement between data and model acceptance. Remaining differences
  are at the few percent level.

  \begin{figure*}
    \centering
    \includegraphics[width=16cm,draft=false]{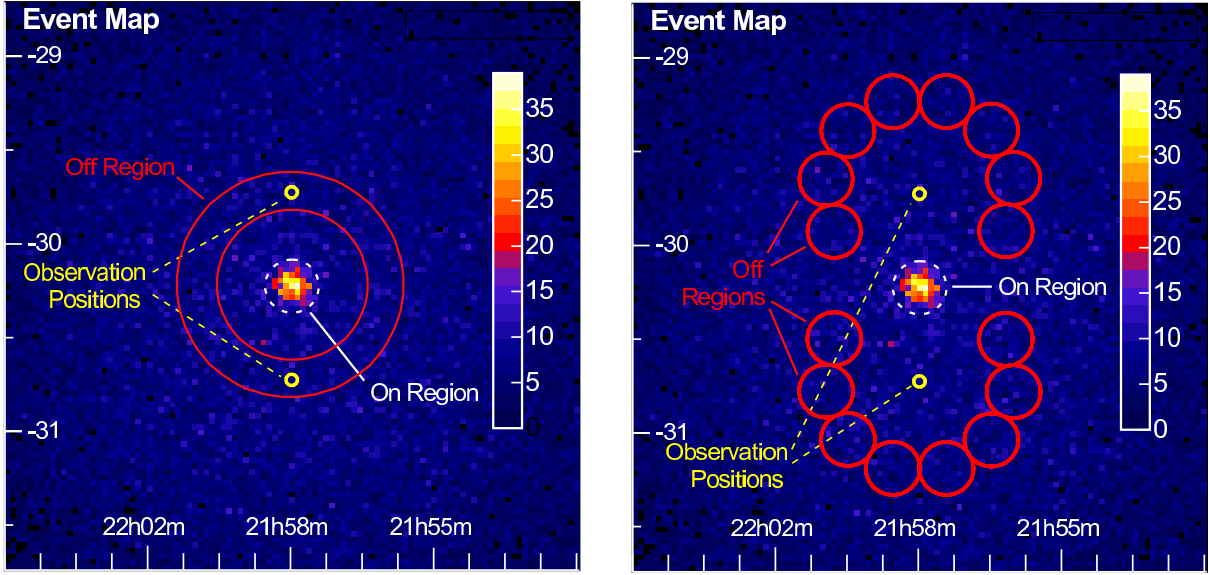}
    \caption{Count map of \gr-like events from 5~hours of
      H.E.S.S. observations of the active galaxy
      PKS~2155--304~\citep{HESSPKS2155}. Note that the data were taken
      in wobble mode around the target position with alternating
      offsets of $\pm 0.5\degr$ in declination. The \emph{ring}-
      (\textbf{left}) and \emph{reflected-region}- (\textbf{right})
      background models are illustrated schematically.}
    \label{schematic}
  \end{figure*}
  \begin{figure*}
    \centering
    \includegraphics[width=16cm,draft=false]{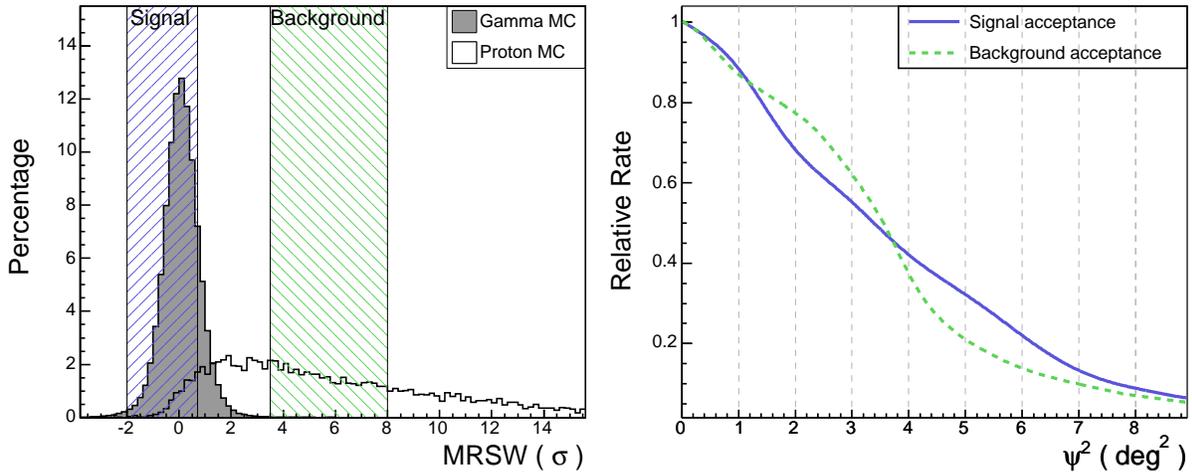}
    \caption{Illustration of the \emph{template}-background
      model. \textbf{Left:} Distribution of the mean reduced scaled
      width (\emph{MRSW}) from \gr\ as compared to proton
      simulations. This parameter is defined as the deviation in units
      of standard deviation of the measured image width from the
      simulated expectation value, averaged over all telescopes
      participating in an event. The separation potential of the MRSW
      is clearly seen, it is frequently used for background
      suppression in \hess\ analyses~\citep{HESSPKS2155}. Events
      falling into the \emph{Signal} region are considered \gr-like
      events, events falling into the background region ($3.5 \sigma
      \le \mrm{MRSW} \le 8 \sigma$) are considered cosmic-ray-like
      events and are used for background estimation. \textbf{Right:}
      System acceptances for the \emph{Signal} and 
      \emph{Background} regimes as indicated in the legend. The
      distributions are generated from \emph{OFF} runs. The
      \emph{Background} acceptance is normalised to the area of the
      \emph{Signal} acceptance in the central 1.5\dg.}
    \label{sketch_template}
  \end{figure*}

  Having discussed the system acceptance function,
  which is crucial for all background models, we now return to
  the principal task of a background model: to provide estimates of
  $N_{\mathrm{off}}$ and
  $\alpha$~(Eq.~(\ref{eq:excess})~\&~(\ref{eq:alpha})).
  Various background estimation techniques are described below.

  \subsection{Ring Background}
  A method that is robust in the face of linear gradients in arbitrary
  directions is the \emph{ring}-background model.  In this model a
  ring around a trial source position (in celestial coordinates) is
  used to provide a background estimate. This is applicable to any
  point in the FoV. The parameter $\alpha$ is approximately
  the ratio of the solid angle of the ring (of typical radius
  0.5$^{\circ}$) to the trial source region
  $\Omega_{\mathrm{on}}/\Omega_{\mathrm{off}}$, and is typically
  chosen to be $\sim$1/7. However, within the ring the acceptance can
  not be assumed to be constant, since the ring covers areas with
  different offsets from the observation position. Therefore an
  acceptance correction function must be used in the determination of
  the normalisation $\alpha$ for each position on the ring. The
  \emph{ring}-background method is illustrated schematically in
  Fig.~\ref{schematic}~(left).

  \subsection{Reflected-Region Background}
  The \emph{reflected-region}-background model was originally
  developed for \emph{wobble} observations~\citep{HEGRACasA,HESSCrab},
  but can be applied to any part of the FoV displaced from the
  observation position. For each trial source position a ring of
  $n_{\mathrm{off}}$ \emph{OFF} regions is used (see
  Fig.~\ref{schematic}~(right)). Each \emph{OFF} region is the same
  size and shape as the \emph{ON} region and has equal offset to the
  observation position (note that here the ring is centred on the
  observation position, while for the ring background technique the
  ring is centred on the trial source position). The method is called
  \emph{reflected-region} method because the \emph{ON} region is
  reflected with respect to the FoV centre to obtain one \emph{OFF}
  region. In the general case as many reflected \emph{OFF} regions as
  possible are then fit into the ring whilst avoiding the area close
  to the trial position to prevent contamination of the background
  estimate by mis-reconstructed \grs. Due to the equal offset of
  \emph{ON} and \emph{OFF} regions from the pointing direction of the
  system, no radial acceptance correction is required with this method
  and $\alpha$ is just $1/n_{\mathrm{off}}$. This is particularly
  helpful for spectral analysis where an energy-dependent radial
  acceptance function would otherwise be required. We note that in
  case the \gr\ source was observed under a large range of offset
  angles with respect to the system pointing direction, for example as
  part of a sky survey, the normalisation $\alpha$ might differ
  substantially from run to run. In this case, a suitable averaging
  procedure has to be applied to both nominator and denominator of
  Eq.~\ref{eq:alpha}: the exposure measure is weighted by a factor
  taking account of the offset of the source from the pointing
  direction (this factor might be calculated as the ratio of the \gr\
  acceptance at the offset of the run to the acceptance at a reference
  offset).

  \subsection{Template Background}
  \begin{figure*}
    \centering
    \includegraphics[width=17cm,draft=false]{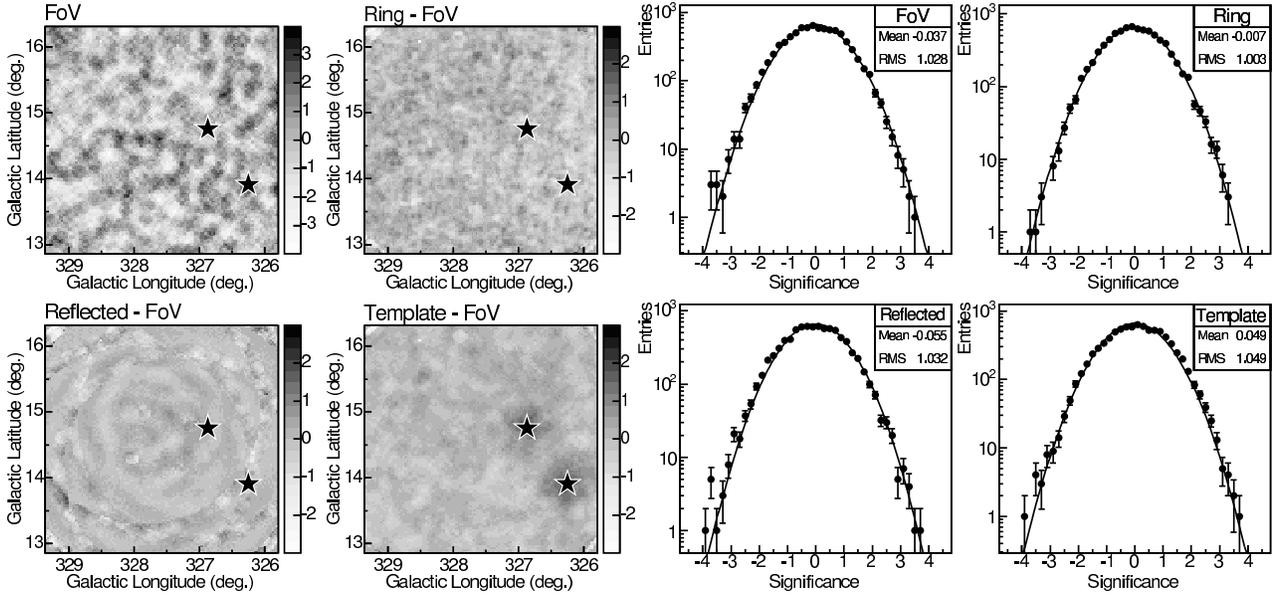}
    \caption{\textbf{Left:} Maps of statistical significance for the
      field around the supernova remnant SN~1006 derived using four
      different background models. 6.3 hours of 4-telescope \hess\
      data without significant \gr\ signal are used. The black stars
      in each field mark the position of bright ($m_{B} = 2.5$ and
      $m_{B} = 2.9$) stars. The upper left-hand plot is produced
      applying the \emph{field-of-view} method, for the purpose of
      comparison, the other plots show the \emph{difference} between
      alternative models and the \emph{field-of-view} model. Note that
      the bins as plotted here are correlated since the significance
      is calculated integrating events in a circle of $0.1\degr$
      radius of each trial source position. \textbf{Right:}
      Distributions of significance in the FoV around SN~1006 for the
      different models. The solid black curve illustrates the expected
      normal Gaussian distribution. As can be seen, deviations from
      the expected behaviour are at the less than 1\% level.}
    \label{significance_maps}
  \end{figure*}

  The \emph{template}-background model was first developed for the
  HEGRA instrument and is described in~\citet{TEMPLATE}. This method
  uses background events displaced in image-shape parameter space
  rather than in angular space. A subset of events failing \gr\
  selection cuts are taken as indicative of the local background
  level. The approach is demonstrated in Fig.~\ref{sketch_template}
  (left). Events falling into the \emph{Background} regime are taken
  as \emph{OFF} counts, \gr-like events from the \emph{Signal} regime
  are \emph{ON} counts. The normalisation $\alpha$ is calculated as
  the number of events in the \emph{Signal} regime, excluding the
  source region, divided by the number of events in the
  \emph{Background} regime. A correction factor depending on the
  position in the FoV has to be applied to $\alpha$ since the system
  responds differently to the cosmic-ray-like than to the \gr-like
  events. Therefore, an additional radial acceptance curve for the
  \emph{Background} regime has to be determined. This cosmic-ray
  acceptance curve depends on the choice of \emph{Background}
  regime. In practice it turns out that the system acceptance becomes
  very different from the \gr\ acceptance if \emph{Signal} and
  \emph{Background} regime are too far apart. This is undesirable
  because the necessary correction factor would vary strongly within a
  FoV, potentially increasing systematic uncertainties. The choice of
  \emph{Background} regime is thus a compromise between good
  separation from the \emph{Signal} regime and small $\alpha$ (i.e.\
  reasonable event statistics), and
  obtaining a background system acceptance function which does not
  differ substantially from the  \gr\  acceptance.
  For the particular choice of \emph{Background}
  regime applied here, the difference of the two system acceptance
  curves can be seen in Fig.~\ref{sketch_template} (right). In the
  central part, for event directions close to the system pointing
  direction, the two curves are very similar. For offset angles larger
  than $\sim 1.5\dg$ pronounced differences occur. The ratio of the
  \gr\ and cosmic-ray acceptances, which is required to determine
  $\alpha$, will not be constant over the FoV.

  The \emph{template} method has the advantage that the background is
  determined in the same region as the signal and hence any localised
  problem, for example due to a bright star, will affect both signal
  and background. Whether the effect is equal for both and therefore
  cancels out depends on the choice of the \emph{Background}
  regime and has to be checked from case to case. A drawback of this
  method is that exact knowledge, not only of the \emph{Signal}, but
  also the \emph{Background} acceptance is required,
  potentially increasing systematic uncertainties related to the
  modelling of the system acceptance.

  Another method that has been applied to H.E.S.S. data is the
  \emph{weighting} method which is related to the \emph{template}
  background. Signal and background are estimated simultaneously from
  the same portion of the sky. Events with directions associated with
  a certain sky bin are assigned two weights, one for the assumption
  that it is a signal event, one that it is a background
  event. Subtraction of the accumulated bin content yields the
  \gr\ excess. This approach is not pursued further, but is
  described in detail elsewhere~\citep{MARIANNE}.

  \subsection{Field-of-View Background}
  For the \emph{field-of-view}-background model, the entire field
  (excluding regions of known \gr\ emission) is used for the
  normalisation of an acceptance model to the data and the
  normalisation $\alpha$ approaches zero. The acceptance model is
  derived from the set of \emph{OFF} runs mentioned above. Given an
  observation at a certain zenith angle, a model background map is
  created by rotating the radial acceptance curve (cf.\
  Fig.~\ref{acc_shape}) of the corresponding zenith angle band. The
  advantage of this model is that it can be readily applied to
  extended sources and results in the highest possible statistical
  significance. However, the method is sensitive to deviations of the
  true system acceptance from the model applied.

  \begin{figure*}
    \centering
    \includegraphics[width=14cm,draft=false]{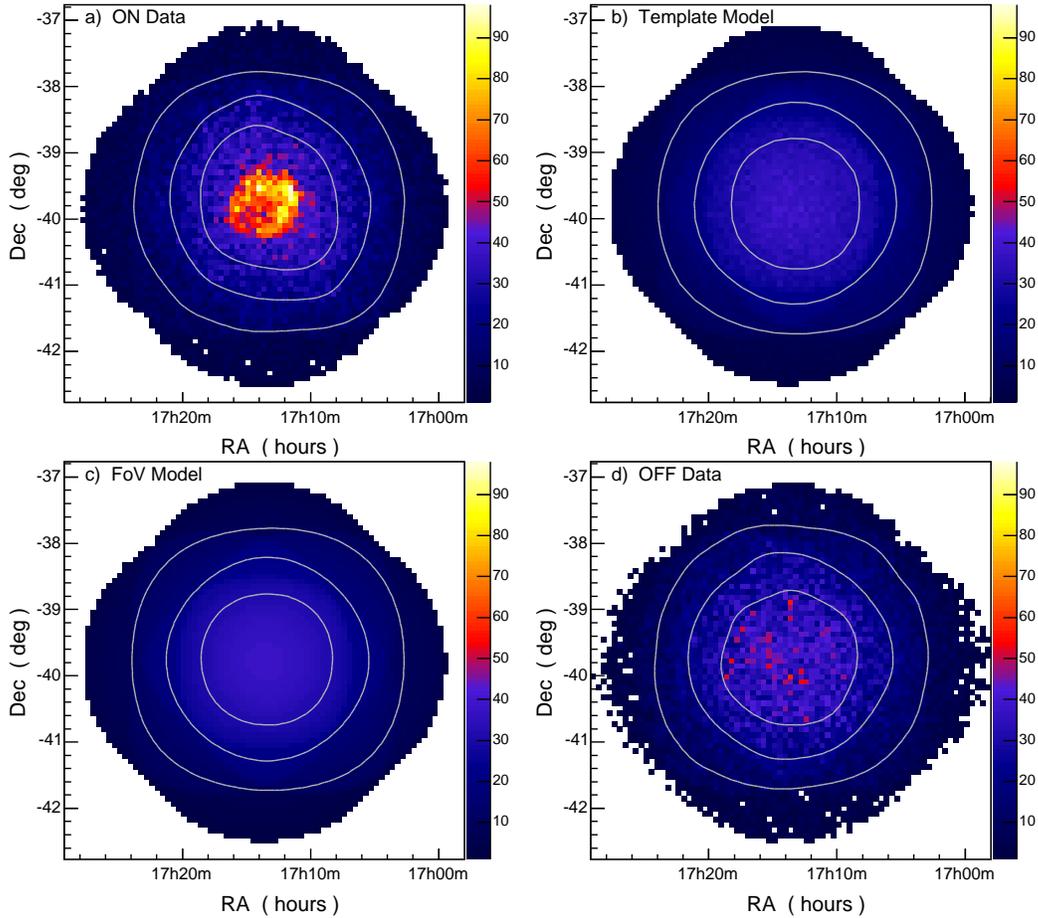}
    \caption{Illustration of different background models for the case
      of the supernova remnant RX~J1713.7$-$3946. Note that
      \emph{hard} cuts have been applied here. \textbf{a)} Raw \gr\
      count map generated from \hess\ 2004 data for the field around
      the remnant. \textbf{b)}~-~\textbf{d)} Normalised background
      maps derived using three different approaches (discussed in the
      main text). Overlaid on all four images are white contours for
      illustration. They are equally spaced at 10, 20, and 30 counts
      and are deduced from a Gaussian-smoothed version of the raw
      colour images to reduce the impact of statistical fluctuations.}
    \label{skymaps_onoff}
  \end{figure*}

  \subsection{Classical ON/OFF Background}
  As mentioned earlier, in traditional \emph{ON/OFF} mode twice the
  observing time is required for each source, providing a strong
  disincentive for this approach to background
  modelling. Nevertheless, the \emph{ON/OFF} mode has a powerful
  advantage in that no assumption is made for the system acceptance,
  except that it is the same in both exposures. As the \emph{ON} and
  \emph{OFF} runs have identical pointing direction in the horizon
  system, the only assumption made is that the acceptance is not
  dependent on conditions fixed in celestial coordinates, such as
  stars and night-sky background light. This advantage motivates a modified form of \emph{ON/OFF} analysis
  that has been applied to \hess\ data. A sizable fraction of the
  fields observed with the \hess\ instrument contain no significant
  \gr\ signal. These data can be used as an archive of \emph{OFF}
  data. For a given set of \emph{ON} runs, a set of \emph{OFF} runs
  matching in zenith angle is selected from the archive. The
  normalisation, $\alpha$, between \emph{ON} and \emph{OFF} runs is
  deduced from the total event numbers in the two runs,
  \emph{excluding} the nominal \gr\ source region.

  \begin{figure*}
    \centering
    \includegraphics[width=17cm,draft=false]{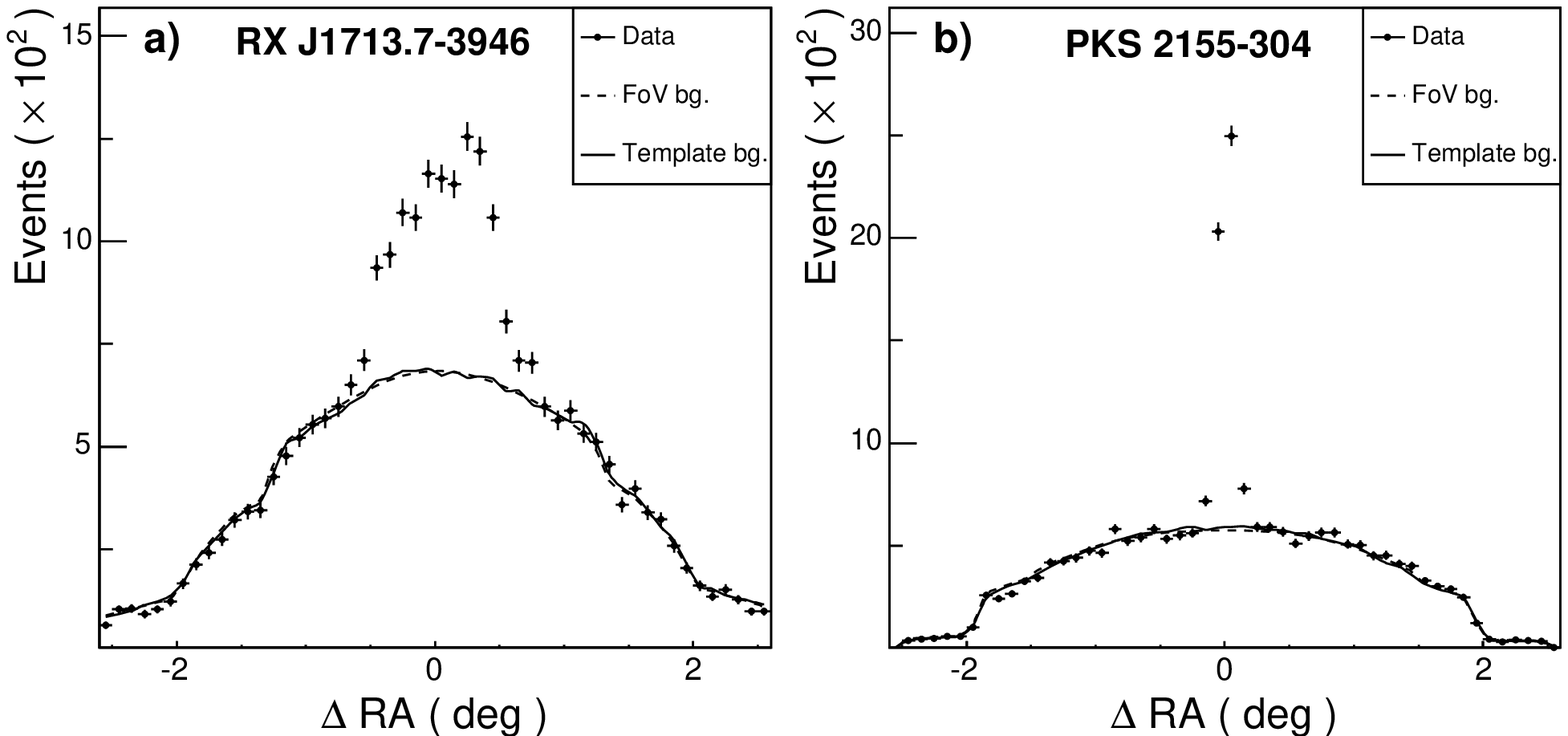}
    \caption{Illustration of the agreement between data points and
      background model (solid and dashed lines). \hess\ data for the
      extended \gr\ source RX\,J1713.7--3946 (\textbf{a)}) are
      compared to PKS~2155$-$304, a point-like extra-galactic \gr\
      source for which a high-statistics data set
      exists~\citep{HESSPKS2155}~(\textbf{b)}). Slices along Right
      Ascension (RA) through the centre of (and fully encompassing)
      the sources, are shown. Two background models are shown, the
      \textit{field-of-view}- (dashed line) and the
      \textit{template}-background model (solid line). Note that the
      steps in the distributions are artifacts of the analysis: for
      each observation run, the usable range of the FoV is restricted,
      and the distributions are produced from different observation
      positions.}
    \label{slices}
  \end{figure*}

  \section{Model Comparison}
  \label{sec:comparison}
  A satisfactory background model must meet one main criterion: when
  applied to many trial positions it should produce a normal
  distribution of significance for an empty field. To test this, four
  of the background models described here have been applied to the
  \hess\ data of SN~1006, which (as already mentioned) shows no
  evidence for \gr\
  emission~\citep{HESSSN1006}. Figure~\ref{significance_maps}~(left)
  compares significance maps of this one field derived using four
  different background models. \emph{Hard} cuts have been applied and
  the significance at each trial sky position was calculated
  integrating \emph{ON} events within a circle of $0.1\degr$ radius.
  A map of absolute significance is shown only for the
  \emph{field-of-view} model. To ease comparison, for other models the
  difference in significance to the reference
  \emph{field-of-view}-model is plotted. The maps show satisfactory
  agreement with each other on the $1~\sigma$-level, within the
  expected statistical fluctuations. The difference maps are roughly
  constant throughout the field, with the only exception being the
  regions close to two bright stars in the lower right of the FoV. As
  is described in more detail below, bright stars cause a reduction in
  the local rate of events, producing a dip in all significance maps
  except that derived using the \emph{template}-background model, in
  which case the bright stars influence also the background
  estimate. For the particular choice of background regime employed
  here, the \emph{template} model slightly over-corrects for the dip
  in the \gr\ acceptance, producing a positive significance at the
  star positions.

  Figure~\ref{significance_maps}~(right) shows the distributions of
  significance values of each trial source position for the maps shown
  in Fig.~\ref{significance_maps}~(left). The regions close to the two
  bright stars have been excluded. The distributions show satisfactory
  agreement with the expected normal Gaussian. We note that deviations
  on the mean significance slightly larger than
  $1/\sqrt{N_{\mathrm{trials}}}$ are expected due to the correlations
  between the signal and background estimates in neighbouring
  positions. These distributions show that, at least for this FoV, the
  systematic error on the assignment of a statistical significance to
  the signal at a given position is at the $<0.1 \sigma$
  level. Furthermore, the agreement between different models suggests
  that under normal circumstances they all provide valid background
  estimates with fluctuations at or close to the expected Poisson
  behaviour.
  
  Figure~\ref{skymaps_onoff} compares 
  a \gr\ count
  map of the field around the supernova remnant RX~J1713.7$-$3946 to
  three different background model maps. The count map is generated
  from \hess\ data from 2004~\citep{HESSRXJ1713_II} with four
  observation positions, all offset by 0.7\dg\ from the centre of the
  remnant. In each observation run, the usable range of the FoV is
  restricted to the central 2\dg\ around the observation position
  causing edges in acceptance when overlaying data from different
  observations. The background models are normalised by $\alpha$ and
  shown on the same scale. It is apparent that the different models
  have different levels of statistical fluctuations. For the
  \emph{template}-background map the statistics are reasonably good,
  $\alpha \approx 1 / 14$ when choosing a background regime of $3.5
  \sigma \le \mrm{MRSW} \le 8 \sigma$ (cf.\
  Fig.~\ref{sketch_template}) and \emph{hard} cuts. $\alpha$ is
  practically zero in case of the \emph{field-of-view}-background
  model, statistical fluctuations are negligible, consequently the
  background map is very smooth. The \emph{OFF}-data map (from the
  \emph{ON/OFF} analysis) has the largest~$\alpha$~$(\approx 1$), and
  statistical fluctuations are at a considerable level resulting in a
  comparatively low statistical significance of the signal. For
  illustration contour lines are overlaid on all four sky images in
  the figure. Apart from differing event statistics, the three
  background models are in good agreement both in terms of shape and
  absolute level. They clearly provide an appropriate description of
  the background of the \gr\ count map shown in
  Fig.~\ref{skymaps_onoff}~a).

  A systematic comparison of the background level estimated by
  different models was performed for the whole \hess\ 2004
  Galactic plane survey.  Figure~4 of \citet{HESSSCAN2} shows the
  correlation between the background estimate for each grid point in
  the sky derived using \emph{ring}- and \emph{template}-background
  models. The correlation is close to linear over a large dynamic
  range. The spread is consistent with statistical fluctuations in
  $N_{\mathrm{off}}$. The slope of the correlation is 1.007 and
  both background estimates are consistent within 1\% -
  see~\citet{HESSSCAN2} for details.
  
  Another demonstration of the validity of different background models
  is shown in Fig.~\ref{slices}. For two \hess\ data sets with
  significant \gr\ signal, slices along Right Ascension and
  encompassing the sources are shown. Overlaid, in both cases, are the
  \emph{field-of-view} and \emph{template} background models for these
  data sets. In both cases, at different regions in the sky, for an
  extended and a point-like \gr\ source, there is clearly a good match
  between both models and data in regions outside the \gr\
  sources. Moreover, it is evident that the features of the gamma-ray
  morphology of RX~J1713.7--3946 after background subtraction are
  robust and remain unchanged when applying different background
  models. Note that this can also be seen in Fig.~6 of
  \citet{HESSRXJ1713_II} which shows a linear correlation of gamma-ray
  excess counts for the sky region around RX~J1713.7--3946 for two
  different background subtraction methods, the \emph{weighting}
  method (mentioned above and discussed elsewhere~\citep{MARIANNE})
  and the \emph{field-of-view} method.

  \section{Effect of Stars}
    \label{sec:stars}
\begin{figure}
    \resizebox{\hsize}{!}{\includegraphics[draft=false]{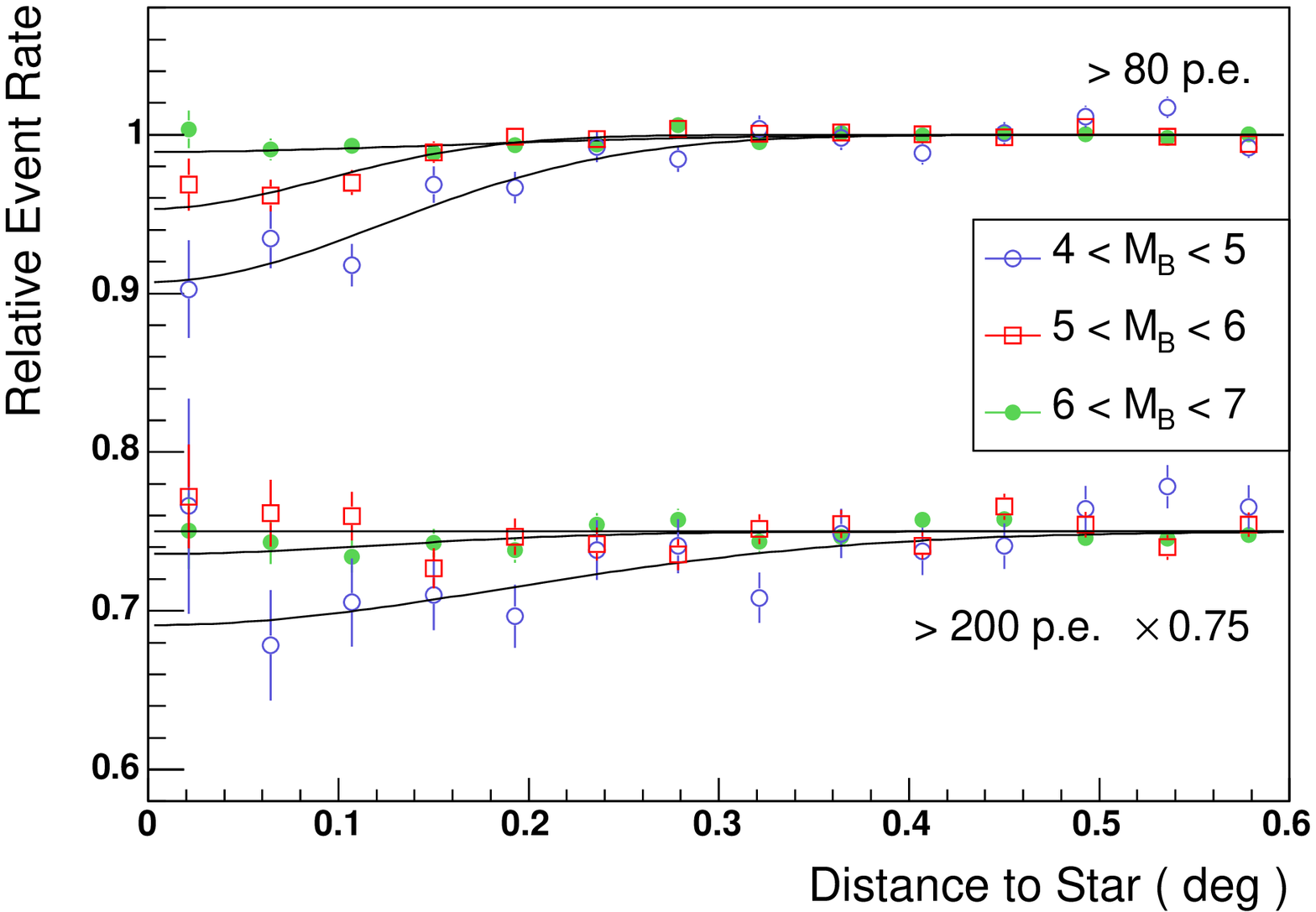}}
    \caption{Effect of bright stars on the event rate of the system
      for standard and hard cuts. The relative event rate is shown as
      a function of the distance to bright stars for different
      B-magnitude bands. The 200~p.e.\ size-cut data set has been
      scaled by 0.75 for clarity.}
    \label{star_dips}
  \end{figure}

  All background estimates presented here rely on the homogeneity of
  the (\gr\ or hadron) acceptance across the FoV. While detector
  acceptance inhomogeneities are typically of the order of 3\% or
  less, they may reach higher values in special cases such as large
  zenith-angle observations or in the presence of strong
  sky-brightness variations (most frequently due to
  stars). Figure~\ref{star_dips} shows as an example the event rate in
  arbitrary units, as a function of the distance to bright stars in the
  FoV, for different bands of stellar B-band magnitude. The curves
  were derived from the H.E.S.S.\ Galactic Plane survey dataset,
  averaging over all stars in the respective magnitude interval. It
  can be seen that for stars with B-magnitude smaller
  than 5, the event rate at the position of the star decreases
  noticeably. This effect can be explained as a consequence of the
  automatic switching off of pixels when DC illumination reaches
  potentially damaging levels. For events in which the shower core is
  located close to one of the telescopes, the event for that telescope
  will have a hole in the middle of its Cherenkov image due to pixels
  being switched off by the star. An image with a hole in the middle
  might a) be thrown away by the image cleaning, or b) fail the shape
  cuts. Therefore less events are reconstructed in the direction of
  the star. The effect increases with increasing brightness of the
  star. The histograms shown in Fig.~\ref{star_dips} are derived using
  \emph{std} cuts. For \emph{hard} cuts the effect is less
  severe. This is expected, as the influence of individual pixels that
  are switched off is smaller for the larger images required with the
  increased image amplitude cut of 200~p.e.. Thus, for the \emph{hard}
  cuts, no significant dip at the position of stars can be seen in the
  magnitude 5 to 6 band. For the magnitude 4 to 5 band, a significant
  dip occurs for both sets of cuts. The magnitude of this effect is
  similar to that found previously for the HEGRA telescope
  array~\citep{GERDHEGRA}.
  
  \section{Choice of Model}
  \label{sec:choice}

  \begin{table*}
    \label{table:bg_properties}
    \centering
    \begin{tabular}{|c||c|c|c|c|c|}
      \hline & \emph{ring} & \emph{reflected-region} & \emph{template} &
      \emph{field-of-view} & \emph{ON/OFF}\\\hline
      Contemporaneous & Y & Y & Y & Y & N\\\hline
      FoV position & N & Y & Y & N & Y\\\hline
      Sky position & N & N & Y & N & N\\\hline
      Event statistics & Y & Y & Y & Y & N\\\hline
      Event type & Y & Y & N & Y & Y\\\hline
    \end{tabular}\\[1.0ex]
    \caption{Overview of the properties of the different background
    models described here. For
    each feature (described on page~\protect\pageref{properties})
    we quote if a given model
    fulfills (\emph{Y}) or fails (\emph{N}) this condition.}
  \end{table*}

  In the following we summarise the advantages and disadvantages of
  the available background models and list the analysis types they are
  most suited for. We start by listing properties of an \emph{ideal}
  background model:

  \renewcommand{\labelenumi}{\alph{enumi})}
  \label{properties}
  \begin{description}
    \item[\textbf{Contemporaneous:}] Background events for a given
      data set should stem from a contemporaneous observation period
      to avoid incompatibilities due to ageing effects of the telescope
      system.
    \item[\textbf{FoV position:}] Background events should be
      accumulated at the same or a similar position in the FoV,
      meaning that the angular distance to the system pointing 
      direction should be equal to the signal events (because of the
      angular dependence of the system acceptance, cf.\
      Fig.~\ref{acc_shape}).
    \item[\textbf{Sky position:}] To assure a similar level of
      night-sky background light background events should be recorded
      from the same region of sky.
    \item[\textbf{Event statistics:}] To reduce fluctuations, the
      event statistics of the background should be considerably larger
      than the signal one, implying a normalisation factor $\alpha \ll
      1$.
    \item[\textbf{Event type:}] Signal and background events should be
      of the same type, from the same region in image-parameter phase
      space to assure a similar system response and reduce the
      importance of the correct system-acceptance model.
  \end{description}
  It is obviously impossible to fulfil all of these requirements with
  a single background model. Any choice can only be a compromise. It
  is therefore important to apply different models to a data set
  thereby cross-checking the results. Table~\ref{table:bg_properties}
  classifies the models discussed in Section~\ref{sec:models} in terms
  of the properties listed above. The advantages and shortcomings of
  the different models are:

  \begin{description}
  \item[\textbf{\emph{ring} background:}]

    The model has the advantage of providing a conceptually simple
    prescription for the background determination. It is rather
    insensitive to deviations of the actual relative acceptance of the
    data set from the model acceptance function, as it only relies on
    the relative normalisation in a limited nearby area around the
    source bin in the FoV. Any linear gradients in the system
    acceptance are averaged out because of the summation on the
    ring. However, when testing larger source extensions, larger ring
    radii are required, demanding better accuracy for the relative
    acceptance correction across a larger portion of the FoV. For the
    determination of energy spectra, this method is disfavoured, since
    the acceptance curve changes with energy (as shown in
    Fig.~\ref{acc_energy}~(left)). Any attempt to correct for this,
    e.g.\ by determining acceptance curves in energy bands, would
    introduce another source of systematic uncertainty. Additionally,
    the \emph{OFF} events have a different distribution of offsets
    from the centre of the FoV than the \emph{ON} events. Since the
    effective areas used for the spectral analysis depend on the
    camera offset, the different offset distribution introduces again
    an additional systematic uncertainty, even if one corrects for
    it. We note furthermore that in the case of several \gr\ sources
    or indications for sources in the FoV, a case which is becoming
    more common in Galactic observations with the current generation
    of experiments, the \emph{ring} background suffers from additional
    systematic uncertainties, that is, possible \gr\ contamination: If
    the source is surrounded by several $3\sigma$ spots, some of which
    might be actual so far undetected sources, they would all be
    included in the background estimate and lead to a systematic
    over-subtraction. 

  \item[\textbf{\emph{reflected-region} background:}]

    It has the advantage that it is independent of the exact shape of
    the acceptance function. It simply relies on the assumption of
    radial symmetry of the acceptance. Additionally the distribution
    of offsets from the centre in the \emph{ON} and \emph{OFF} events
    is the same. This makes the model especially suited for the
    background estimation for energy spectra. However, this approach
    relies on a suitable observation strategy, it cannot be applied if
    the observation positions of a data set are within an extended
    source region, or, as mentioned above, in case of too many other
    \gr\ sources in the FoV. In this case one either ends up in a
    situation where it is not possible to define a reflected
    background region without overlap with an actual known source
    region, or, in case of close-by sources just below detection
    limit, one might obtain a \gr\ contaminated background estimate.

  \item[\textbf{\emph{template} background:}] 

    This technique has the advantage that it is better suited to
    largely extended sources (which fill a sizable fraction of the
    experiment's FoV) than the \emph{ring}-background method as long
    as an acceptance model (say from \emph{OFF} data) is available. It
    is, however, sensitive to uncertainties in the relative acceptance
    determination between the \emph{Signal} and \emph{Background}
    regime across the FoV. Large differences of the system--acceptance
    functions in the two regimes potentially increase systematic
    uncertainties of the normalisation factor $\alpha$. 

    For the estimation of energy spectra the background estimate must
    consist of events with a similar distribution of estimated
    energies to the background events of the source region. The
    \emph{template} model does not meet this criterion since the
    events in the background regime will differ in estimated energy
    from those in the signal regime. Also the problem of the
    acceptance curve changing with energy is present as in the case of
    the \emph{ring} background but more severely since it applies to
    the signal as well as to the background regime.

  \item[\textbf{\emph{field-of-view} background:}] 

    The model can readily be applied to any data set to investigate
    the source morphology. It is especially well suited for very
    extended sources that cover a large fraction of the FoV and yields
    the maximum possible signal significance since the normalisation
    factor $\alpha$ is practically zero. The caveat is that it is
    sensitive to deviations of the model from the true system
    acceptance. For example, pronounced night-sky brightness
    variations within a single field and data sets with unbalanced
    \emph{wobble} observation offsets might cause distortions of the
    true system acceptance.

  \item[\textbf{\emph{ON/OFF} background:}] 

    This classical approach is a robust method to perform cross-checks
    and explore systematic uncertainties of spectra of very extended
    sources. Its advantage is that it can be applied to any data set,
    independent of the source size and the observation strategy
    pursued for a given source (as opposed to the
    \emph{reflected-region} approach, which relies on observation
    positions outside the source region). Its caveats are the loss of
    a factor of two in \emph{ON}-source observation time and possible
    changes in the night-sky background level between \emph{ON} and
    \emph{OFF} data.

  \end{description}

  To summarise, the best suited background estimation technique for
  the extraction of various aspects of the \gr\ signal are: 
  \begin{itemize}
  \item[$\Box$] \emph{Source detection}: the \emph{ring}-background
    model has in general fewest systematic biases for this
    purpose. Only in the case of busy sky fields which contain
    multiple potential gamma-ray sources issues due to gamma-ray
    contamination arise.
  \item[$\Box$] \emph{Spectral analysis}: a \emph{reflected-region}
    background is most suitable due to the identical offset
    distribution of \emph{ON} and \emph{OFF} regions. For very
    extended sources, however, or for extended sources which have not
    been observed with sufficiently large \emph{wobble} offset, or in
    case there are other \gr\ sources in the FoV and one cannot define
    background regions, the \emph{ON/OFF}-background model is the only
    appropriate one.
  \item[$\Box$] \emph{Morphology of very extended sources}: the
    \emph{field-of-view} model provides an effective way to
    investigate the morphology of sources too large to be effectively
    handled by other methods.
  \end{itemize}
  
  \subsection{Optimal Wobble Offset}
  \begin{figure}
    \resizebox{\hsize}{!}{\includegraphics[draft=false]{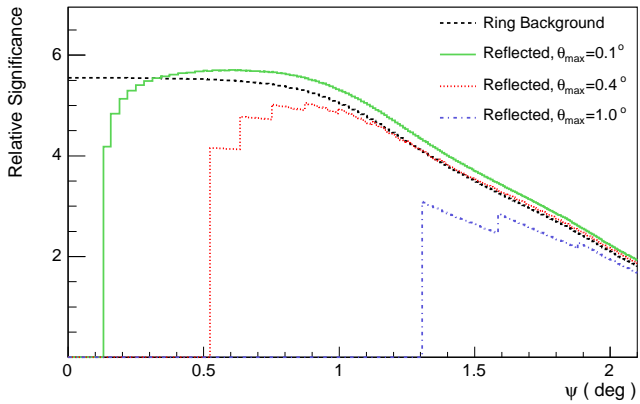}}
    \caption{The impact of an offset between observation position and
      target source (\emph{wobble} offset) on the source significance.
      For a \emph{reflected-region} background an offset is required
      to provide \emph{OFF} regions at equal offset. The relative
      significance has a broad plateau between 0.5 and 0.7 degrees for
      point-like sources.}
    \label{opti_wobble}
  \end{figure}
  
  Besides choosing a background model best suited for a given source
  and the analysis task, the optimal observation strategy needs to be
  considered. It is governed by the source properties and the
  preferred background model to be applied. To derive a background
  estimate and extract a signal from a single dataset it is normally
  necessary to observe a potential source with an offset with respect
  to the pointing direction of the system.  For a given background
  model and for known radial system acceptance, the optimum
  observation (or \emph{wobble}) offset $\Theta_{\mathrm{opt}}$ can be
  calculated. This offset is defined as the one which maximises the
  significance per unit observation time ($S/\sqrt{t}$) of the source.
  For \emph{ring} and \emph{template} backgrounds
  $\Theta_{\mathrm{opt}}=0$, but this is undesirable for two
  reasons. The first is that only if the source is offset from the
  system pointing direction can the one-dimensional radial system
  acceptance be extracted from the data set under study (because of
  the exclusion of the source region for acceptance generation). The
  second reason is that a spectral analysis becomes more difficult and
  one introduces systematic difficulties for extended sources since
  the \emph{reflected-region} method cannot be applied. In fact, for
  the \emph{reflected-region} background, $\Theta_{\mathrm{opt}}$ is a
  compromise between the number of available \emph{OFF} regions (which
  increases with increasing offset) and the fall-off of the system
  acceptance for large offsets. In Fig.~\ref{opti_wobble} we explore
  the significance as function of wobble offset. For point sources in
  \hess, $S/\sqrt{t}$ exhibits a rather flat plateau between $0.4\dg$
  and $0.7\dg$. For moderately extended sources
  ($\sigma_{\mathrm{source}}\sim0.2^{\circ}$) the optimal offset
  increases to $0.7\dg$-$1.0\dg$.  For very large sources the best
  strategy is to observe just outside of the source, with some safety
  margin to account for the finite angular resolution, so that one
  reflected off region is available without \gr\ contamination.
  Targeted \hess\ observations are typically taken at $0.5\dg$ or
  $0.7\dg$ offsets.
  
  \section{Conclusions}
  
  Several different background models are available for ground based
  Cherenkov astronomy. Different models are appropriate for different
  purposes. Searches for weak sources are best performed with the
  robust \emph{ring}-background model. For spectral analysis the
  \emph{reflected-region} background is favoured. Extracting the
  morphology of extended sources is often most reliable with the
  \emph{field-of-view} or \emph{template} models. In general, a
  comparison of several models (with different systematics) is
  necessary to establish the existence of a new source. It is
  important to remember that the estimated statistical significance of
  a source is largely irrelevant if background systematics are not
  under control.
  
  \section*{Acknowledgements}
  The authors would like to acknowledge the support of their host
  institutions, and additionally support from the German Ministry for
  Education and Research (BMBF). Specifically, JH acknowledges the
  support of the BMBF through Verbundforschung Astro-Teilchenphysik
  (05CH5VH1/0).  We would like to thank the whole H.E.S.S.\
  collaboration for their support, especially Werner Hofmann, Gerd
  P\"uhlhofer, Gavin Rowell and Christian Stegmann for their useful
  remarks, and Richard White for his careful reading of the
  manuscript.

  \label{lastpage}
  
  \bibliographystyle{aa}

\end{document}